# A tunable multi-color "rainbow" filter for improved stress and dislocation density field mapping in polycrystals using x-ray Laue microdiffraction

Odile Robach[a*], Jean-Sébastien Micha[b], Olivier Ulrich[a], Olivier Geaymond[c], Olivier Sicardy[d], Jürgen Härtwig[e], François Rieutord[a]

[a]CEA-Grenoble, INAC / SP2M / NRS, [b]UMR SPrAM CNRS-CEA-UJF, CEA-Grenoble, INAC, [c]CEA-LITEN, MINATEC Campus, [a,b,c]17 rue des Martyrs, 38054 Grenoble Cedex 9, France, [c]Institut Néel, CNRS 25 rue des Martyrs, F-38042 Grenoble, France, [d]ESRF, 6, rue Jules Horowitz, BP 220, 38043 Grenoble Cedex 9, France. E-mail: odile.robach@cea.fr

**Synopsis**

A technique to measure the energy profiles of Laue spots in x-ray Laue microdiffraction is presented. It uses a single-crystal diamond filter that attenuates several well-defined energies in the incident white-beam spectrum. A first application to lattice parameter measurements is demonstrated.

**Abstract**

White beam x-ray Laue microdiffraction allows fast mapping of crystal orientation and strain fields in polycrystals, with a submicron spatial resolution in two dimensions. In the well crystallized parts of the grains, the analysis of Laue spot positions provides the local deviatoric strain tensor. The hydrostatic part of the strain tensor may also be obtained, at the cost of a longer measuring time, by measuring the energy profiles of the Laue spots using a variable-energy monochromatic beam. A new "Rainbow" method is presented, which allows measuring the energy profiles of the Laue spots while remaining in the white-beam mode. It offers mostly the same information as the latter monochromatic method, but with two advantages : i) the simultaneous measurement of the energy profiles and the Laue pattern; ii) the rapid access to energy profiles of a larger number of spots, for equivalent scans on the angle of the optical element. The method proceeds in the opposite way compared to a monochromator-based method, by simultaneously removing several sharp energy bands from the incident beam, instead of selecting a single one. It uses a diamond single crystal placed upstream of the sample. Each Laue diffraction by diamond lattice planes attenuates the corresponding energy in the incident spectrum. By rotating the crystal, the filtered-out energies can be varied in a controlled manner, allowing one to determine the extinction energies of several Laue spots of the studied sample. The energies filtered-out by the diamond crystal are obtained by measuring its Laue pattern with an other 2D detector, at each rotation step. This article demonstrates the feasibility of the method, and its validation through the measurement of a known lattice parameter.

## 1. Introduction

Synchrotron radiation x-ray Laue microdiffraction using white beam has been used for more than a decade (Chung & Ice, 1999, MacDowell et al., 2001, Kunz et al., 2009, Ice & Barabash, 2007, Ice & Pang, 2009, Ulrich et al., 2011, Maaß et al., 2006, Ice et al., 2004, Tamura et al., 2000, 2002, 2003, Larson et al., 2002, Kirchlechner et al., 2010, 2011, Hofmann et al., 2010) to determine the strain and orientation fields in polycrystalline materials, with a submicron spatial resolution, attempting to elucidate the relations between microstructure and mechanical properties. The existing instruments at ALS, APS and ESRF[1] all offer the possibility to switch to monochromatic beam, to measure, via the photon energy, the interplanar distance $d_{hkl}$ of a given ($hkl$) spot.

When the local crystalline quality is sufficient (misorientations / mosaic < 1 mrad in the probe volume), the $d_{hkl}$ measurement may be combined to the Laue pattern measurement to retrieve the complete set of the 6 lattice parameters, and deduce the full elastic strain tensor of the unit cell. This combination requires to maintain the unit cell shape and orientation with respect to the incident beam perfectly constant between the two measurements. For the monochromatic method, this implies to re-position the beam on the inhomogeneous sample with an accuracy better than the typical length inside the sample corresponding to a variation of $10^{-4}$ on the orientation or the strain. The difficulty of this alignment led to develop a white beam method (Robach et al., 2011) to simultaneously measure the Laue pattern on the 2D detector, and the energy of one spot using an energy-resolved point detector mounted sideward on two translation stages. This method remains slow for raster sample scan as the positioning of the point detector depends on the grain orientation : a prior analysis of the Laue pattern is necessary. This analysis is also necessary for the monochromatic method (unless long energy scans are used to measure several peak energies) to set the monochromator energy close to the approximate spot energy before scanning.

In the case of larger micro-misorientations, the shapes of the Laue spots, and the spot displacements associated to probe volume displacement (orientation gradients), may be analyzed to estimate the density of unpaired dislocations (Geometrically Necessary Dislocations (GND's)). The energy width of the spots provides the total dislocation density ρ, independently of the paired or unpaired character of the dislocations (Barabash & Ice, 2012) (with a $\sqrt{\rho}$ dependence when GND's are randomly arranged).

This article describes the first tests of another method for measuring the energy position and width of the Laue spots, based on the concept of a rotating "multi-color filter" : instead of using an incident beam with a single energy (as in the monochromatic mode), a white beam is used, in which several well-defined energies are missing. A similar method has already been proposed for neutrons (Marmeggi, 1984), except that the rotation was applied to the sample and not to the filter.

## 2. Experimental details

### 2.1. Setup and samples

The experiments were performed using the Laue microdiffraction setup of the CRG-IF BM32 beamline at ESRF (Ulrich et al., 2011). A schematics of the experimental setup is shown in Figure

---

[1] Advanced Light Source (Berkeley, USA), Advanced Photon Source (Argonne, USA), European Synchrotron Radiation Facility (Grenoble, France)

1. The standard instrument features micro-focusing optics, a xyz translation stage for the sample holder inclined by 40 degrees with respect to the white incident beam (energy range 5-22 keV) and a 2D detector (#1) above the sample. Upstream of the focusing optics a multi-color filter system was added which includes the following elements : a vertical-gap slit to reduce the beam size down to 0.3x0.3 mm$^2$, followed by a horizontal translation stage to bring the filter in and out of the beam. This stage carries a vertical translation stage, holding a motorized rotation stage (angle $\theta_f$) with a horizontal axis nearly perpendicular to the incident beam (within 4-5 degrees), itself holding a single-crystalline thin diamond plate (the filter). The diamond plate of 3×8 mm$^2$, with (110) orientation and 300 μm thickness makes an angle of approximately 45 degrees with respect to the incident beam. This orientation allows having two of the most intense diamond diffraction lines (the (111)'s) in the 9.5-12.5 keV range. A second 2D detector (#2) is placed upstream of the filter near $2\theta$ = 120 degrees to collect the Laue patterns. This allows calculating the energies of all the beams diffracted by the diamond crystal at any crystal angle, thereby providing the list of energies that will be attenuated in the beam coming to the sample.

The diamond crystal was first mapped (installed in the sample position) by usual Laue microdiffraction, in order to check the absence of deviatoric strain of the unit cell (< $2.10^{-4}$) and the homogeneity of the unit cell orientation (better than 0.2 mrad). It was then installed on the rotation stage, on the path of the incident beam. Scans in filter angle $\theta_f$ over 2.5 or 5 degrees with a 0.0025 degrees step were then performed, while recording on detector #1 the Laue patterns of the sample. Two samples were tested in a first campaign : a Germanium (111) single crystal wafer, and a polycrystalline bi-layer based of yttria-doped zirconia, forming the electrolyte and anode of a half - solid oxide fuel cell (SOFC). The electrolyte layer in this sample consists of grains of a few microns (Villanova et al. , 2010, 2011).

Three other single crystalline Ge samples (numbered #1, #2, #3) with different orientations were also tested in a second campaign, in order to estimate the uncertainty on the lattice parameter. Table 1 summarizes the geometry of the rotating filter and the sample orientation for the various Ge samples. The orientation of the incident beam with respect to the crystal axes of the filter was chosen to be of low symmetry to be far away from degenerate conditions (when several diamond diffracted beams have the same energy). This limits the occurrence of very closely-spaced dips, which are more difficult to analyze, in the intensity vs. $\theta_f$ curves of the sample's Laue spots.

The effect of the filter insertion on the x-ray beam size was characterized and found negligible for the part of the spectrum above 9.7 keV. Figure 2 shows the profiles obtained by scanning a rectangular thin film of gold with well defined edges in front of the microbeam and measuring its fluorescence. The slope of the profile stays constant when inserting the filter.

**2.2. Data analysis**

Each dip observed in the intensity vs. $\theta_f$ curve provides, via its position $\theta_{f\_dip}$, the energy $E_{exp}(hkl_{sample})$ of the corresponding Laue spot. The experimental value of the lattice spacing $a$ can then be obtained by combining $E_{exp}$ with the local grain orientation and deviatoric strain ($b/a$, $c/a$, $\alpha$, $\beta$, $\gamma$) deduced from the Laue pattern. The theoretical value $E_{theor}$ for the spot energy is calculated for a hypothetical lattice parameter $a_0$, then the $dE/E = (E_{exp}/E_{theor}-1)$ provides the deviation $-(a-a_0)/a_0$ on the lattice spacing. When using the unstrained lattice parameter as $a_0$, this directly gives the hydrostatic part of the strain.

Alternatively, one set of six lattice parameters (and the full strain tensor) may be obtained by combining the measured energies and Bragg angles of six well chosen Laue spots whose hkl's are already known. This may be useful when the accuracy on deviatoric strain provided by the Laue pattern is poor (e.g. when Laue spots have a shape incompatible with a precise locating (within 0.1 pixel) of their center of mass).

One important procedure in the analysis is the dip indexation, i.e. the assignment of a $hkl_{filter}$ triplet to a given dip observed on a given $hkl_{sample}$ spot. This step is currently done manually using a $\theta_f$ vs. energy graph derived by analyzing the Laue patterns of the rotating diamond crystal. In this graph, the experimental dips are reported as points (using $a=a_0$ to calculate the $E_{theor}(hkl_{sample})$ values). The $E_{filter}(hkl_{filter}, \theta_f)$ curves of the intense hkl lines of the diamond are also reported. Figure 3a shows for example the graph used to index the dips observed for Ge sample #0. The crossings between the vertical lines $E(hkl_{Ge})$ and the inclined lines $E(hkl_{diamond}, \theta_f)$ provide a list of theoretical dips, and also their $\theta_f$ positions. Each Laue spot from the sample may therefore undergo several extinctions, if the corresponding vertical line crosses several inclined lines. More than 200 filter lines were expected between 5 and 22 keV, but only 64 are shown here, the ones for which $(f_{polarisation} \times f_{atomic})^2$ is larger than 0.1% of the intensity of the most intense line (for which $f_{polarisation}$ is nearly 1). The slope of the $E_{filter}$ curves varies with $hkl_{filter}$, evidencing that different diamond lines provide different energy resolutions for a given angular step. 25 of the diamond lines created detectable dips in this example.

For the Ge samples, the $\theta_f$ positions of the dips were analyzed to retrieve the Ge lattice parameter. Unit cell lengths $a_0$ of 5.6575 Å and 3.5668 Å were assumed for unstrained Ge and diamond respectively. The $a$ values were derived by combining $E_{dip}$ measurements with $E_{theor}$ values deduced from the Laue patterns. They are given as deviations with respect to the theoretical lattice parameter : $dE/E = -da/a = -d\lambda/\lambda$. An important remark is that here the length unit is the diamond lattice parameter. The experimental $(a_{sample}-a_{0\_sample})/a_{0\_sample}$ may therefore contain an unknown $(a_{diamond}-a_{0\_diamond})/a_{0\_diamond}$. One way of checking the diamond lattice parameter would be to use a diamond diffracted beam whose energy coincides with a fluorescence line of the screen of detector #2.

Based on Bragg's law (i.e. neglecting dynamical diffraction effects), the position in $\theta_f$ and in $E_{filter}=E_{Ge}$ of an indexed dip can be calculated from fourteen scalar parameters, eight of which are fixed : the $hkl_{Ge}$, the $hkl_{filter}$, and the diamond and Ge lattice parameters. Six parameters are variable : first the two angles of the incident beam (unit vector $\mathbf{u_i}$) with respect to the Ge unit cell ($\mathbf{u_{i\_Ge}}$), and secondly the geometry of the rotating diamond, which provides the two angles of the incident beam with respect to the diamond unit cell ($\mathbf{u_{i\_dia}}(\theta_f)$) for any filter angle $\theta_f$. This geometry is fully describes by four parameters : the value of $\mathbf{u_{i\_dia}}$ for a given $\theta_f = \theta_{f0}$, and the two angles of the rotation axis $\mathbf{axis_{dia}}$. Here $\theta_{f0}$ is taken at the center of the scan. To include dynamical diffraction effects, other parameters describing the shape of the diamond (e.g. thickness and hkl's of the two faces) would need to be added.

When the geometry of the rotating filter is known via its four parameters, $\mathbf{u_{i\_dia}}$ can be calculated for any $\theta_f$. Each experimental $\theta_{f\_dip}$ can then readily be converted into a $E_{filter}$. Alternatively, a series of $\mathbf{u_{i\_dia}}$ values may be determined by using the refinement of the diamond Laue patterns collected for a series of $\theta_f$ values. $\theta_{f\_dip}$ can then be converted into $E_{filter}$ by interpolating the resulting $E(hkl_{filter}, \theta_f)$ table at $\theta_f = \theta_{f\_dip}$, as shown in Figure 3b.

In practice, the $E(hkl_{filter}, \theta_f)$ tables provided by the Laue patterns allowed us to correctly index a large number of dips, but led to large deviations between the Ge lattice spacing values measured using different dips. This was due to a poor accuracy on $\mathbf{u_{i\_dia}}$. For each pattern, 8 parameters had been refined using the spot positions : the diamond orientation (3 parameters) and the geometry of detector #2 (5 parameters)[2]. The poor accuracy came from the irregular and elongated shape of the diamond spots on detector #2, which led to large mean pixel deviations (around 1.2 to 1.5) between theory and experiment after refinement. A comparatively much better accuracy was available for $\mathbf{u_{i\_Ge}}$, with mean pixel deviations around 0.1 after refinement of the Ge Laue pattern on detector #1.

It was therefore decided to use the equality of the $a_{Ge}$ values derived from the various Ge dips as a criterion to refine the geometry of the rotating filter, taking advantage of the large number of dips available for the Ge (with often several dips per spot). *In fine*, the geometry of the rotating filter was calibrated in two stages. First the diamond Laue patterns collected at the two extreme values of $\theta_f$ provided a guess on the geometry. This allowed the indexation of a large fraction of the Ge dips. Then $\mathbf{u_{i\_dia}}(\theta_{f0})$ and $\mathbf{axis_{dia}}$ were refined to minimize the deviation between the $a_{Ge}$ values obtained from different Ge dips.

## 3. Results

### 3.1. Germanium single crystals

Figure 4a shows the intensity vs. $\theta_f$ profiles for the 32 Laue spots of the Ge single crystal #0 (over a total of 86 spots) that presented one or several measurable extinctions over the scanned 5-degrees range. The Laue patterns of the Ge sample and the diamond filter are shown in Fig. 4b and 4c. The observed dips in Fig. 4a varied between 5 and 50%. The angular width of the dips varied between 0.0025 and 0.05 degrees in $\theta_f$, illustrating the dependence of energy resolution on $hkl_{filter}$. Several dips showed complex shapes, asymetric or with a "S"-shape. Further work is needed to investigate if dynamical diffraction effects in the thick diamond may explain these shapes. A few expected extinctions (not shown) even led to peaks instead of dips (possibly due to diamond-induced changes in the polarization of the incident beam). A substantial number of extinctions were therefore available (here 68) for sample lattice parameter measurements.

Table 2 summarizes the results obtained for the Ge lattice parameter for the various samples, after locating the dips in the intensity vs. $\theta_f$ curves (similar to Fig. 4a), indexing the dips using a $\theta_f$ vs. energy graph (similar to Fig. 3a), and calculating the $dE/E$ for each dip. The number of experimental dips, for a fixed geometry of the rotating filter (and a fixed scanning range), varied from 27 to 60 depending on the sample orientation (for samples #1 to #3). It increased to 68 by (mostly) increasing $\theta_{f0}$ by 2 degrees (sample #0).

When the filter geometry was obtained from the diamond Laue patterns (sample #0, first line), the mean $dE/E$ was large ($-9.8 \times 10^{-4}$), and the agreement between the $dE/E$ values of the different dips was poor (standard deviation $(dE/E)_{std}$ of +/- $38 \times 10^{-4}$). After optimizing $\mathbf{u_{i\_dia}}(\theta_{f0})$ and $\mathbf{axis_{dia}}$ to minimize the $(dE/E)_{std}$ (sample #0, second line), a much better agreement was obtained ($(dE/E)_{std}$ of +/- $1.0 \times 10^{-4}$), with a low mean $dE/E$ (- 0.6 $\times 10^{-4}$), i.e. the measured parameter was in much better agreement with the literature value. This is encouraging, in view of the very simple procedure used for locating the dips (taking the point of minimum intensity in the curve drawn using the most intense pixel of the spot).

Minor corrections (below 15 mrad) on the two angles of $\mathbf{u_{i\_dia}}(\theta_{f0})$ were sufficient, while larger corrections were necessary for the two angles of $\mathbf{axis_{dia}}$ (50 to 300 mrad). This reflects the weak sensitivity of $\mathbf{u_{i\_dia}}(\theta_f)$ with respect to the direction of the rotation axis for the small $\theta_f$ range used here (+/- 2.5 degrees).

The comparison between samples #1, #2 and #3 allowed an investigation of the accuracy of the refined filter geometry. The geometry was first refined separately on each sample. Then the geometry refined on sample #1 was used to calculate the $dE/E$'s for samples #2 and #3. For the sample-per-sample optimization, the mean $dE/E$ varied between $-2.6 \times 10^{-4}$ and $+0.7 \times 10^{-4}$. For the optimization on sample #1, it varied between $-0.9 \times 10^{-4}$ and $+1.9 \times 10^{-4}$. The error on the mean $dE/E$ that comes from the uncertainty on the geometry of the rotating filter is therefore below +/-2 $\times 10^{-4}$.

Differences in accuracy were noticeable between samples. The final $(dE/E)_{std}$ was 3 to 5 times larger for samples #1, #2 and #3 than for sample #0. One possible explanation is a degradation of the energy resolution between the two series of measurements. The micro-focusing optics was indeed changed between the two campaigns, to provide a smaller beam size, and the new optics accepted a larger area of the incident beam, leading to twice larger beam divergences on both the sample and the filter. Keeping a low beam divergence may therefore be an issue. The $(dE/E)_{std}$ after refinement also varied with the sample orientation (between 3.2 and 5.0 for samples #1, #2 and #3) for a given filter geometry. These variations may be related to variations in the proportion of sharp dips in the data set.

The analysis performed here is rather crude, and the obtained deviation on the $dE/E$ should be perfectible by :

- using a proper description for the shape of the dips, taking into account the dynamical diffraction effects occurring in the thick diamond crystal.
- using the integrated intensity of the Ge spots instead of the intensity of a single pixel.

The direct determination of the geometry of the rotating filter using the Laue patterns on detector #2 may also become more accurate by using a more realistic hypothesis on the shape of the diamond spots (to describe the elongation due to dynamical diffraction effects (Yan & Noyan, 2005)). It is not clear yet if it may reach the accuracy already obtained through the criterion of equalizing the $dE/E$'s of all the dips of a Ge single crystal. Another option would be to apply the monochromatic method to the rotating diamond filter to further refine its geometry.

As an element of comparison with the monochromatic method, in terms of number of measured spot energies, the same diamond used as a monochromator set on the (-111) (see red line on Fig. 3a) would provide only 10 measurements of the $da/a$ for an equivalent angular scan. With a Si(111) monochromator, a 1000 points angular scan between 10 and 11 keV provided the energy of 6 Ge spots. The two methods are however not fully comparable, as several spot energies are measured more than once with the "rainbow" method.

### 3.2. Solid Oxide Fuel Cell sample with micron-sized grains

The next test consisted in checking the sensitivity of the method for a sample with micron-sized grains. Figure 5 shows the one-pixel intensity profiles vs. $\theta_f$ (cf. Fig. 4a) for 32 of the 34 spots showing detectable attenuations for the SOFC sample. These spots were among the 173 most intense (out of around 500) of the multi-grain Laue pattern recorded on a single point of the sample. A photograph of the sample surface and the Laue pattern are also shown. Dip depths up to 60% were observed, indicating that the method should also work here. The analysis of the full local strain tensor for the grains in this sample will be described later. Here a single scan caused extinctions on spots from several grains, providing the *da/a* for each of them. By comparison, the monochromatic method with large scans (e.g. 1 keV) also probes several grains, but without the constant check on beam position provided by the simultaneous Laue pattern measurement. The white beam method with the energy-resolved detector probes only one grain for each detector position.

---

[2] Here "geometry of the detector" means the orientation of the incident beam and the 3D position of the unique point in the sample supposedly at the origin of all diffracted rays. All these parameters are referenced to the Oxyz frame attached to the detector.

## 4. Conclusion and perspectives

The "rainbow" method provides fast energy profiles for numerous Laue spots, with simultaneous collection of the Laue pattern. The large number of independent measurements of the lattice parameter (68 for a Ge sample with a 5 degrees scan of the filter) should improve the accuracy when trying to measure hydrostatic strain with respect to a known unstrained lattice parameter. A first attempt at measuring the known lattice parameter of an unstrained bulk crystal (Ge) led to a mean *da/a* of $0.6 \cdot 10^{-4}$ (relative error). The +/- $1.0 \cdot 10^{-4}$ standard deviation obtained here for a set of 68 measurements is certainly perfectible, by fine tuning of the experimental setup and data analysis.

As a tool to study the full local elastic strain tensor, this method should allow :

- improving the reliability of measurements performed by combining the Laue pattern (deviatoric strain) and the energy of one or several spots (absolute value of $d_{hkl}$)

- measuring the full tensor using only spot energies (thanks to the large number of available energy measurements). This may allow to extend the domain of application of stress measurements to probe volumes with higher orientation gradients.

- performing the data collection without prior analysis of the Laue pattern (the simultaneously attenuated energies covering a large energy range). For ultimate strain accuracy, online analysis of Laue pattern will remain useful. This will allow one to adapt the mean $\theta_f$ of the scan in order to maximize the number of crossings between intense sample diffraction lines and intense filter diffracted lines (cf. Fig. 3a). The usefulness of a second diamond rotation stage for a simultaneous optimization of the number of sharp dips remains to be investigated.

Continuous filter scans with a fast-readout detector (e.g. pixel detector) installed near the sample's 2D detector should allow fast mapping of both Laue pattern and energy profile.

Here the method was tested on lattice parameter measurements but it also provides the spot energy width. This should allow fast simultaneous measurements of dislocation densities for both paired and unpaired dislocations. Short angular scans around a single dip should be sufficient, using the deepest and best-resolved dip (e.g. one associated to the $(-111)_{filter}$ line). One application should be the monitoring of the total dislocation density during in situ tensile and compressive mechanical tests of single-crystalline micro-pillars (Kirchlechner et al., 2010, 2011, Maaß et al. 2006), which require fast measurements without sample motion. This should facilitate the monitoring of the first glide system, which is often difficult to detect via spot elongation in the Laue patterns, as it gives few GND's compared to secondary systems.

## Acknowledgements

We thank the CEA-CNRS staff of the BM32 beamline for technical help with the microdiffraction setup, the ESRF for providing the x-ray beam, P. Gergaud, O. Castelnau, G. Renaud, and T. Zhou for participation in the preliminary discussions, H. Isern and P. Montmayeul for technical support, A. Filhol for pointing out the existence of Marmeggi method, and J. Baruchel for the suggestion to use a fluorescence line of the detector screen for calibration. Financial support was provided by the AMOS ANR-10-NANO-015 project led by J.L Rouvière. Laue pattern analysis was performed using the LaueTools software (LaueTools web site). An alternative name of "Batterman filter" was proposed by N. Tamura for the method.

**Figure Captions**

**Figure 1** Experimental setup : the multi-color filter setup (vertical gap slit, *x* translation, *z* translation, $\theta_f$ rotation, 2D detector #2 at $2\theta = 120$ degrees in a vertical diffraction plane) is placed about 1 m upstream of the Laue microdiffraction setup (H and V gap slits, Kirkpatrick-Baez mirrors for micro-focusing, *xyz* sample translation stage, sample at 40 degrees, 2D detector #1 at $2\theta = 90$ degrees in a vertical diffraction plane). The filter creates numerous well-defined dips in the energy spectrum of the white incident beam. These dips shift in energy with the rotation, and successively attenuate the various Laue spots of the sample.

**Figure 2** Effect of the insertion of the filter on the size of the x-ray micro-beam. Fluorescence profiles on a thin rectangular gold layer on silicon, without (black symbols) and with (red symbols) the filter. The *x* translation is horizontal and perpendicular to the incident beam. The *y* translation is at 40 degrees of the incident beam, in the vertical plane containing the beam. Beam size on sample (*x,y*) : (0.8, 1.7) µm.

**Figure 3** (a) : the energy vs. filter angle graph used for indexing the dips in the case of Ge sample #0. The energies of the Ge and diamond diffracted beams were calculated from the Laue patterns. The cross points between the $E_{diamond}$ curves and the $E_{Ge}$ vertical lines provide the location of the theoretical dips. The color code gives a first hint of the dip depth : the theoretical intensity of the diamond diffracted beams (=1.0 for the most intense beam). The circles mark the 68 dips experimentally detected on 32 of the Ge spots. Only Ge beams that present dips are shown. Visual comparison between experimental and theoretical dip positions allows one to index the dips. For each Ge diffracted beam, several vertical lines are needed, as different dips may correspond to different $hkl_{Ge}$ harmonics. (b) interpolation used to convert the experimental $\theta_{f\_dip}$ position into an experimental energy $E_{exp}(hkl_{sample})$, after having indexed the dip using (a).

**Figure 4** Measurements on the Ge single crystal (sample #0) : (a) intensity of 32 of the 86 Laue spots, vs. filter angle (5 degrees scan). For each spot, the intensity of a fixed pixel is plotted (the pixel of maximum intensity at $\theta_f = 0$). Intensities are normalized to their values at $\theta_f = -2.5$ degrees. The curves are shifted vertically for clarity. (b,c) : Laue patterns of the Ge sample (b) (micro-beam, detector #1) and the diamond filter (macro-beam, detector #2, $\theta_f = 0$ (blue) and $\theta_f = 0.5$ degrees (red)).

**Figure 5** Measurements on a polycrystal with micron-sized grains (electrolyte side of a "half" solid-oxide-fuel-cell). The x-ray microbeam is at a fixed position with respect to the sample. (a) intensity of 32 Laue spots from several zirconia grains, vs. filter angle (2.5 degrees scan) (see Figure 4 for details). (b) optical microscope image (field of view : 30x20 µm²). (c) multi-grain Laue pattern : positions of the 173 most intense spots.

**Table 1** Geometry of the rotating filter in the two campaigns, and orientation of the Ge samples, as given vy the hkl's of various characteristic vectors. $\mathbf{u_i}$ is along the incident beam. **axis** is along the rotation axis. $(z_1-y_1)_{Ge}$ gives the hkl's for a (hypothetical) Ge spot at the center of detector #1. This is used to describe the Ge orientation in rotation around the incident beam. "dia" and "Ge" subscripts are used for coordinates with respect to the diamond and Ge lattices respectively.

| sample | $\mathbf{u_{i\_Ge}}$ | $(z_1-y_1)_{Ge}$ | $\mathbf{u_{i\_dia}}(\theta_{f0})$ | $\mathbf{axis_{dia}}$ |
|---|---|---|---|---|
| 0 | -0.79 -0.33 -1 | 0.95 -0.81 1 | 0.98 0.77 1 | 0.24 1 -0.92 |
| 1 | 0.09 0.02 1 | -0.82 -0.83 -1 | -0.92 -0.74 -1 | -0.18 -0.96 1 |
| 2 | 0.38 0.18 1 | -0.91 0.81 -1 | " | " |
| 3 | 0.65 0.46 1 | -1 0.31 -0.57 | " | " |

**Table 2** Lattice parameter *a* measured on four Ge single crystals of various orientations. Values are given as deviations $dE/E = -d\lambda/\lambda = -da/a$ with respect to $a_0 = 5.6575$ Å. Data on sample #0, and on samples #1, #2 and #3, were collected during two different experimental campaigns. The $n_{dip}$ values of $dE/E$ were obtained with a 5-degrees scan in $\theta_f$ (step 0.0025 degrees). The refined geometry of the rotating filter is given as deviations with respect to the initial geometry : $dz(\mathbf{u_i})$ and $dx(\mathbf{u_i})$ for the incident beam (initially along *y*). $dy(\mathbf{axis})$ and $dz(\mathbf{axis})$ for the filter rotation axis (initially close to *x*). "opt" indicates the sample used for the refinement of the filter geometry. For the first entry on sample #0, the geometry of the rotating filter was not optimized : the dip energy was obtained by interpolating at $\theta_f = \theta_{f\_dip}$ the $E(hkl_{filter}, \theta_f)$ table provided by the diamond Laue patterns.

| | dE/E of Ge dips ($10^{-4}$) | | | $n_{dip}$ | opt | angular corrections (0.1 mrad) on $\mathbf{u_{i\_dia}}(\theta_{f0})$ and $\mathbf{axis_{dia}}$ | | | |
|---|---|---|---|---|---|---|---|---|---|
| # | mean | std +/- | range | | | dz ($\mathbf{u_i}$) | dx ($\mathbf{u_i}$) | dy (axis) | dz (axis) |
| 0 | -9.8 | 38 | 238 | 68 | no | 0 | 0 | 0 | 0 |
| 0 | -0.6 | 1.0 | 5.4 | 68 | 0 | 6.0 | -13.8 | 50 | -40 |
| 1 | -0.9 | 5.0 | 23.2 | 60 | 1 | 7.4 | 12.4 | 300 | 50 |
| 2 | 1.9 | 3.9 | 18.2 | 37 | 1 | " | " | " | " |
| 3 | -0.7 | 3.9 | 20.0 | 27 | 1 | " | " | " | " |
| 2 | 0.7 | 3.6 | 17.9 | 37 | 2 | 6.8 | 11.8 | 200 | 20 |
| 3 | -2.6 | 3.2 | 19.8 | 27 | 3 | 6.8 | 10.8 | 200 | 35 |

Fig.1

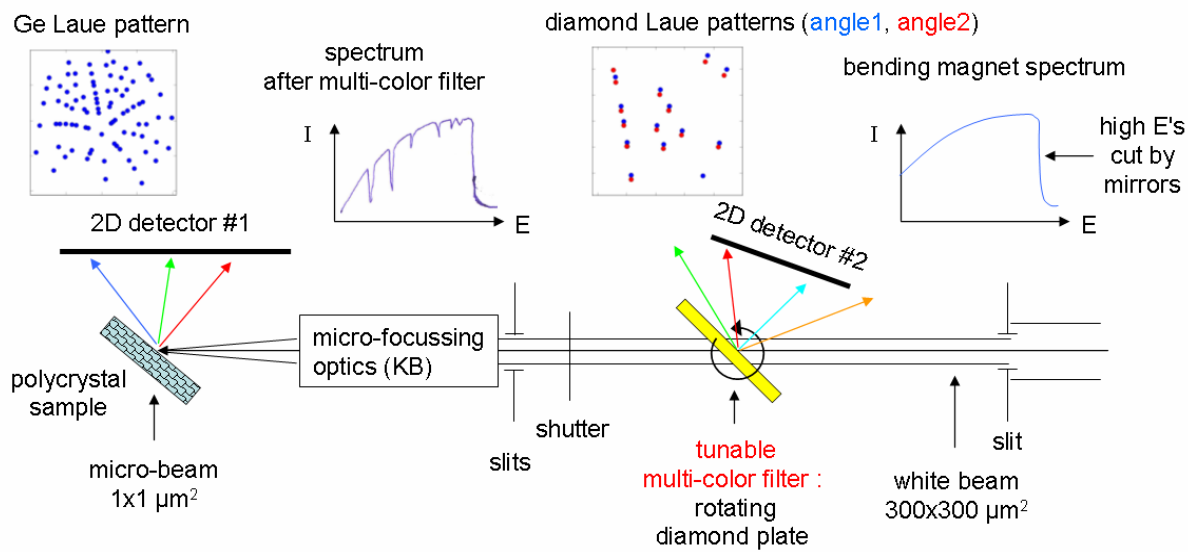

Fig 2 :

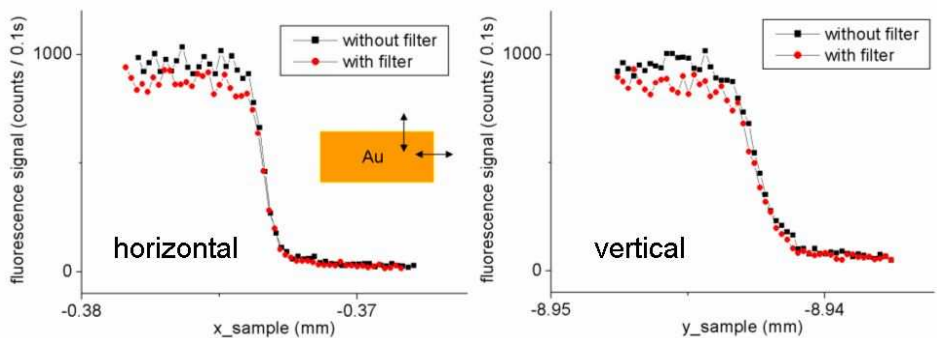

Fig. 3a :

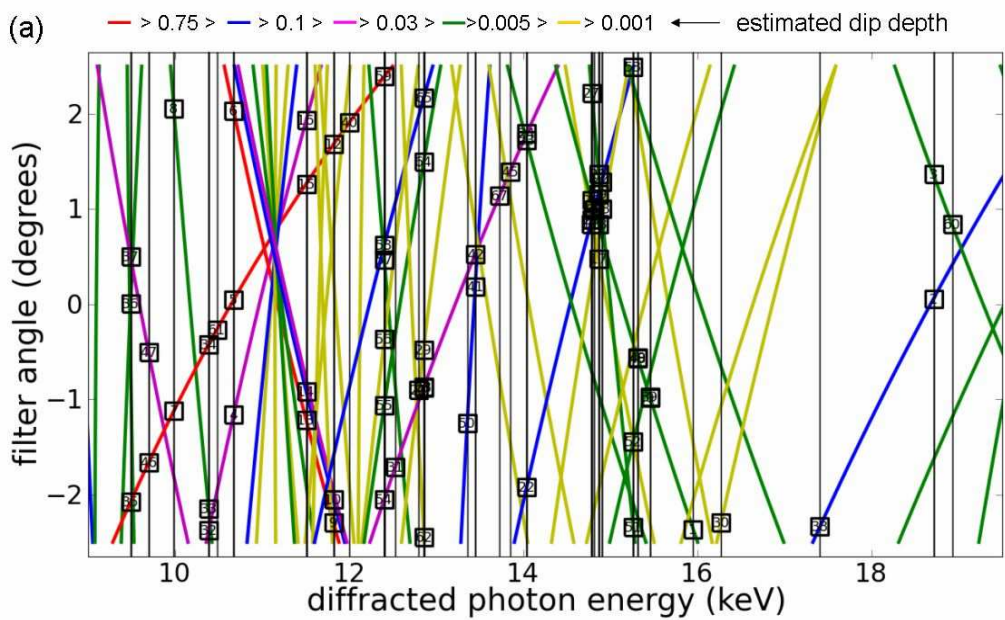

Fig. 3b :

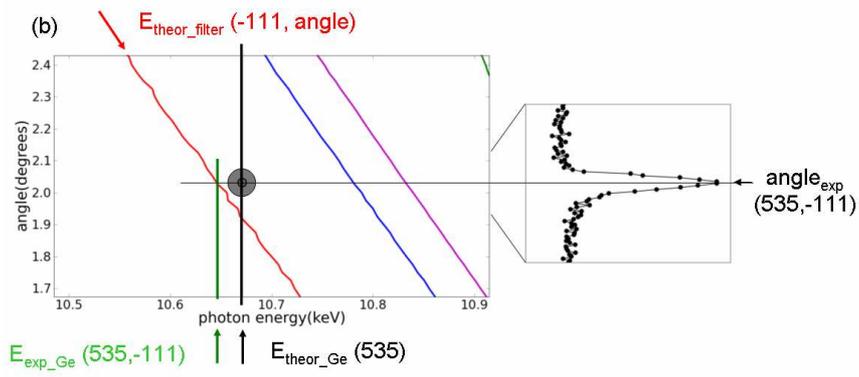

Fig. 4a :

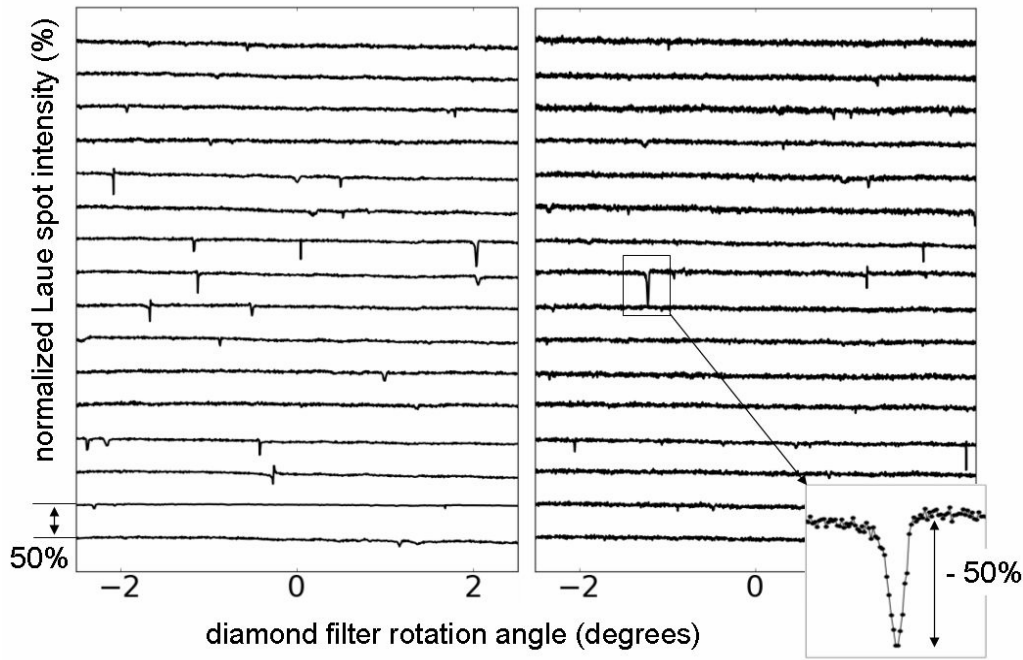

Fig. 4b, c :

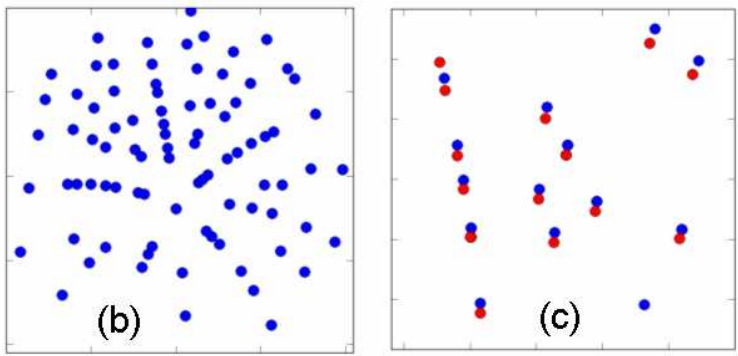

Fig. 5a :

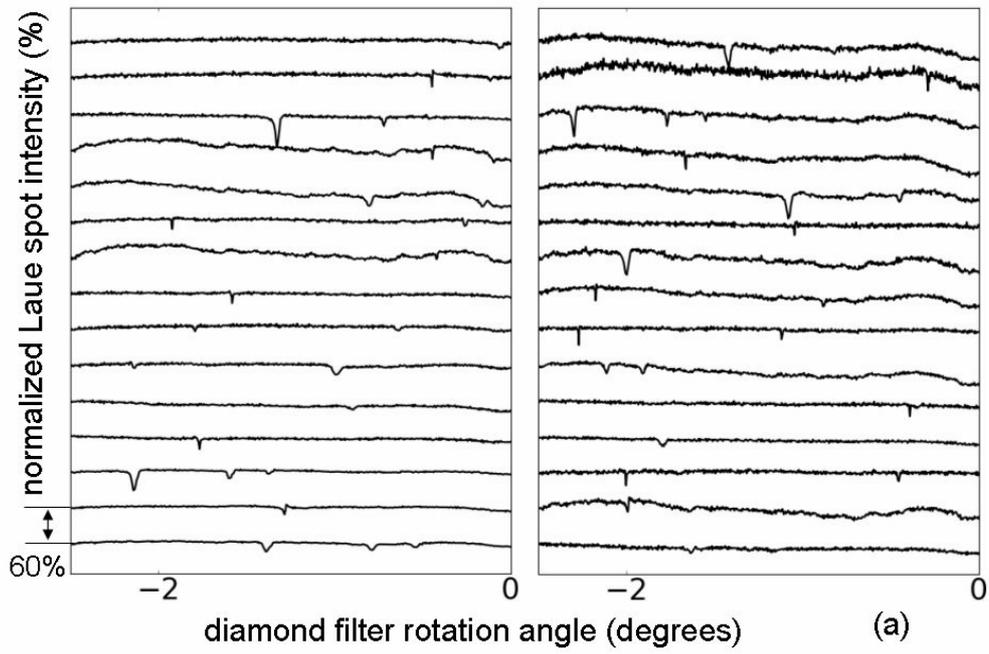

Fig. 5b,c :

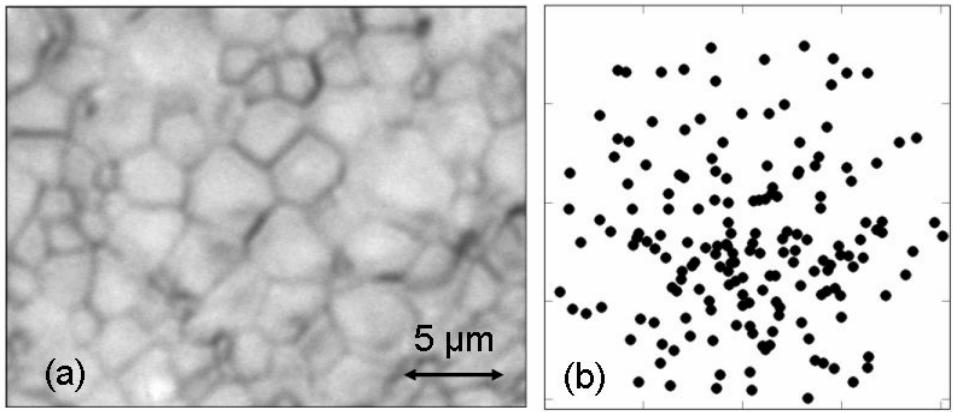